\documentclass[journal]{IEEEtran}
\usepackage{amsmath,amssymb,amsfonts}
\usepackage{algorithm}
\usepackage{algorithmic}
\usepackage{array}
\usepackage[caption=false,font=normalsize,labelfont=sf,textfont=sf]{subfig}
\usepackage{textcomp}
\usepackage{stfloats}
\usepackage{url}
\usepackage{bm}
\usepackage{verbatim}
\usepackage{graphicx}
\usepackage{cite}
\usepackage{booktabs}
\usepackage{xcolor}
\begin{document}

\newcounter{proposition}
\setcounter{proposition}{0}
\newenvironment{proposition}[1][]{\refstepcounter{proposition}\textbf{Proposition \arabic{proposition}:} \rmfamily}{\medskip}

\title{Bayesian Cramér-Rao Bound for Sensing Performance in Meta-Backscatter Systems}

\author{
        Mengyuan~Cao,~\IEEEmembership{Graduate Student Member, IEEE},
        Xu~Liu, ~\IEEEmembership{Graduate Student Member, IEEE},
        and Hongliang~Zhang,~\IEEEmembership{Member, IEEE}.
        \thanks{
        M. Cao, X. Liu and H. Zhang are with School of Electronics, Peking University, Beijing 100871, China (e-mail: caomengyuan@pku.edu.cn, xu.liu@pku.edu.cn, hongliang.zhang@pku.edu.cn). 
	}
}

% The paper headers
% \markboth{Journal of \LaTeX\ Class Files,~Vol.~14, No.~8, August~2021}%
% {Shell \MakeLowercase{\textit{et al.}}: A Sample Article Using IEEEtran.cls for IEEE Journals}

% \IEEEpubid{0000--0000/00\$00.00~\copyright~2021 IEEE}
% Remember, if you use this you must call \IEEEpubidadjcol in the second
% column for its text to clear the IEEEpubid mark.

\maketitle

\begin{abstract}
Meta-backscatter system that utilizes meta-material sensors is a promising enabler for future environmental sensing, offering distinct advantages such as low cost, zero-power consumption, and robustness. Specifically, the electromagnetic response of the sensor, typically characterized by a frequency-selective absorption profile, is affected by the environmental conditions, allowing the estimation of these conditions from the reflected signal. However, it remains unclear what estimation accuracy can be achieved fundamentally. Motivated by this gap, we quantify this accuracy limit using the Bayesian Cramér-Rao bound (BCRB), which provides a lower bound on the mean-squared error for the environmental condition. Establishing this limit is challenging because the electromagnetic response of the sensor is distorted by the channel fading, while the channel estimation is infeasible since the sensors cannot be configured to predefined states to generate training data. To address this challenge, we consider the joint BCRB of the channel coefficient and the environmental condition in a multicarrier framework. The BCRB of the environmental condition is then obtained by selecting the corresponding element from the joint BCRB. An analysis of the derived BCRB reveals the impact of the absorption peak shape and the number of subcarriers. The derivation and analysis of the BCRB are verified through simulations.
\end{abstract}
% the reflection coefficient is distorted by both the wireless channel and the noise. Furthermore, 


\section{Introduction}

\IEEEPARstart{F}{uture} sixth-generation (6G) wireless systems will support a wide range of sensing-centric applications, such as healthcare monitoring and intelligent warehousing, which require pervasive and large-scale sensing capabilities. This anticipated demand imposes stringent requirements on the new generation of Internet-of-things~(IoT) sensors, namely ultra-low cost, minimal power consumption, and long-term maintenance-free. However, traditional sensors struggle to meet these requirements, as they typically rely on active electronic components and dedicated energy supplies to perform sensing and wireless transmission~\cite{b9}.
% To meet this demand, the number of IoT devices is expected to increase by more than 10-fold compared with current 5G deployments~\cite{b8}.

Recently, meta-material sensors have emerged as a promising solution for the above issues\cite{b10}. Composed of sub-wavelength resonator structures printed on a dielectric substrate and integrated with environmentally sensitive materials, the sensor can reflect the incident electromagnetic wave with a characteristic frequency-domain absorption feature. As environmental conditions change, the spectral profile of this absorption feature shifts accordingly. The environment-dependent absorption response is embedded in the end-to-end propagation path and manifests as a multiplicative component of the reflection channel. Therefore, the environmental condition can be inferred from the reflected signals, without requiring active electronic components\cite{b11}.

In the literature, several works considered using meta-material for sensing functions, such as humidity, temperature, and gas concentration\cite{b4,b14,b16}. The authors in \cite{b4} investigated the humidity and concentration sensing performance of three double-sided resonator structures in a waveguide testbed. In~\cite{b14}, a meta-material sensor for various biomolecules was proposed. The proposed sensor consists of a basic single-square split ring resonator (SRR) with an integrated inverted Z-shaped unit cell for dual-band resonance. In \cite{b16}, the authors proposed a meta-backscatter system that leveraged sub-wavelength units to concentrate reflected power and extend communication range. Nevertheless, existing works mainly focus on sensor structure design and system optimization, while the fundamental theoretical limits on the estimation accuracy of meta-backscatter sensing systems remain unexplored.

% A deep-learning-based joint optimization algorithm for sensor design and deployment was proposed in \cite{b5} to achieve high-accuracy temperature and humidity monitoring. 
% b7: proposed a meta-IoT system with inhomogeneous meta-material sensor structures that broaden reflection coverage to support receivers at arbitrary angles. 

% In \cite{b6}, the authors designed a dual-functional meta-IoT network for integrated sensing and communication, and the sensor structures and beamforming scheme are jointly optimized to improve both sensing and communication performance. 

Motivated by this gap, this work considers the meta-backscatter system using meta-material sensors and investigates its fundamental sensing capability limits by deriving a closed-form Bayesian Cramér-Rao Bound~(BCRB) for the environmental condition of interest, which provides a lower bound on the achievable estimation error. Furthermore, we analyze the derived BCRB under a Rician fading channel, assuming a Lorentzian spectral absorption feature. Specifically, we investigate the influence of several system parameters, including the shape of the absorption peak and the number of subcarriers. These analytical results offer fundamental insights into the sensing performance and establish a benchmark for meta-material sensor structure optimization.

The primary challenge stems from the infeasibility of direct channel estimation, as the sensors cannot be configured to predefined states to provide training examples\cite{b18}. Therefore, unlike conventional CRB analyses that assume known channel conditions, our framework must account for the uncertainty of the wireless channel. In response to the above problem, we treat the environmental condition and the channel coefficients as jointly unknown parameters and derive the joint BCRB. The BCRB of the environmental condition is then obtained by selecting the corresponding element from the joint BCRB. Simulations are conducted to validate the derived BCRB.

% The rest of this letter is organized as follows. In Sec. \ref{sec:sys_BCRB}, the system model is provided, and then the BCRB is derived. In Sec. \ref{sec:BCRB_an}, we anlyzed the derived BCRB. Sec. \ref{sec:simu} presents simulation results. Finally, conclusions are drawn in Sec.~\ref{sec:con}.

% In \cite{b3}, the authors proposed a sensor structure with two split ring resonators and a polymer/single walled carbon nanotube composite ink serves as sensitive material for the detection of the temperature and CO2. 

% Several challenges arise for the calculation of BCRB. First, the environment parameter affects the received signal through the reflection coefficient, which is distorted by the wireless channel and additive noise. Both effects must be rigorously incorporated into the performance analysis. Second, since the sensors cannot configure themselves to predefined states and provide training examples, direct channel estimation is infeasible, and the analysis must be conducted under unknown channel conditions. In response to these problems, we treat both the sensing target and the channel coefficients as jointly unknown parameters and derive the joint BCRB. The BCRB of the sensing target is then obtained by selecting the corresponding element from the joint BCRB.

\section{The BCRB for Sensing Target}
\label{sec:sys_BCRB}
In this section, we first establish the system model of the signal, and then the BCRB is derived and analyzed.

\begin{figure}
    \centering
    \includegraphics[width=0.75\linewidth]{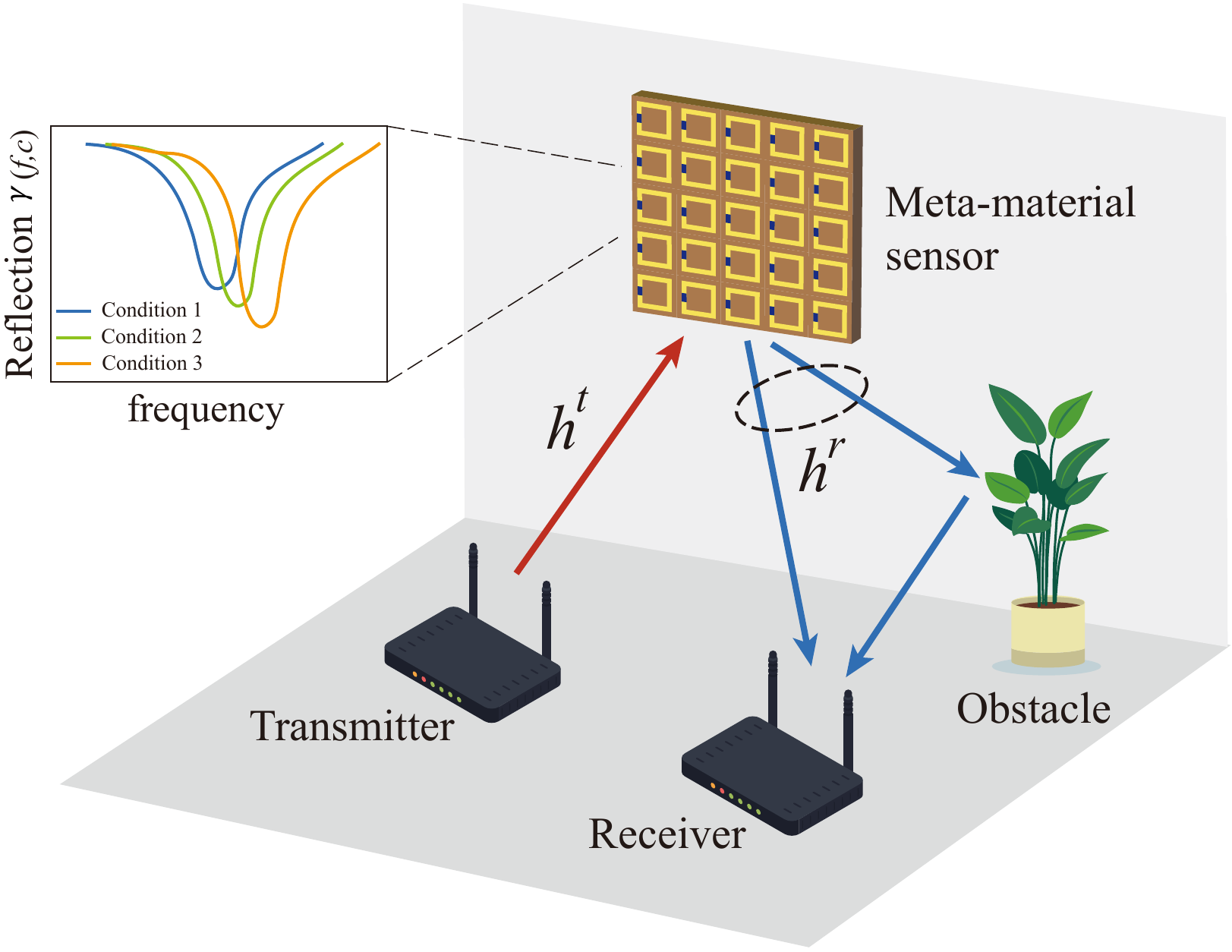}
    \caption{Meta-backscatter sensing system.}
    \label{fig:sce}
\end{figure}

\subsection{System Model}
% \subsection{System Description}
The considered system consists of a pair of transceivers and a meta-material sensor, as illustrated in Fig. \ref{fig:sce}. The transceiver can transmit and receive the signal in a set of discrete subcarriers. The meta-material sensor is a fully passive reflector. The reflection coefficient of the sensor $\gamma(f,c)$ is sensitive to the environmental condition $c$, referred to as the sensing target. As a result, the received signal is modulated by the sensing target, enabling the inference of the sensing target through signal analysis.
% Consequently, the received signal is influenced by the sensing target, and we can obtain the environmental conditions by analyzing the received signals.
The received signal is modeled by
% \begin{align}
%     P(f,c) = 10\log_{10} &\left( \vert H(f,c) \vert ^2  T(f) (\eta_{e} R_W \right. \nonumber\\
%     &\left. + \eta_{s} S(f,c) ) + P_b \right) + e(f),
% \end{align}
\begin{align}
    y(f,c) =  h^r(f)  \gamma(f,c) h^t(f) + n,
    % &= H(f)  \gamma(f,c) + n,
\end{align}
% where $H(f,c)$ is the channel frequency response. $T(f)$ denotes the transmit power. $R_W$ is the reflection coefficient of the wall. $\eta_{e}$ and $\eta_{s}$ are the proportions of the signal incident on the wall and the sensor, respectively. $P_b$ is the bias, which accounts for the influence of ambient environmental signals and measurement error. $e(f)$ is the measurement noise, following Gaussian distribution $\mathcal{N}(0,\sigma^2)$.
where $h^t(f)$ and $h^r(f)$ are the channels from the transmitter to the sensor and from the sensor to the receiver, respectively. $n$ is the noise, following Gaussian distribution $n \sim \mathcal{CN}(0,\sigma^2)$. We collect the received signal on $L$ subcarriers and obtain 
\begin{align}
    y_k = h_{k}^r \gamma_k(c)h^t_k + n_k, \ k = 1,...,L,
\end{align}
where $\gamma_k(c)$ is the refection coefficient for the frequency $f_k$.
% i.e., $\gamma_k(c) = \gamma(f_k,c)$

\subsection{Derivation of the BCRB}

We assume the sensing target $c$ and the channel $h^t$, $h^r$ are all independent random variables. Denote the random parameters as $\bm{\theta} = [c, \bm{h}] \in \mathbb{R}^{1\times (1+4L)}$. Here, $\bm{h}$ collects the channel coefficients over $L$ subcarriers, given by $\bm{h} = [\bm{h}_1, \dots, \bm{h}_L]$. For the $k$-th subcarrier, the channel is given by $\bm{h}_k = [\Re(h^r_k), \Im(h^r_k), \Re(h^t_k), \Im(h^t_k)]$.

% \Re(h^r_1), \Im(h^r_1), \Re(h^t_1), \Im(h^t_1), \dots ,\Re(h^r_L), \Im(h^r_L), \Re(h^t_L), \Im(h^t_L)]
% $\bm{h} = [h^r_1, \dots h_L^r, h_1^t, \dots h_L^t]$.

Based on \cite{b2}, the Bayesian CRB is defined as
\begin{align}
    \text{BCRB} = - \mathbb{E}_{Y,\Theta}\left( \dfrac{\partial^2}{\partial \theta_i \partial \theta_j}\log p(\bm{y}, \bm{\theta})\right)^{-1},
\end{align}
where $\theta_i$ is the $i$-th element in $\bm{\theta}$.
% \begin{align}
%     \text{BCRB} &= - \mathbb{E}_{Y, \Theta}\left(\dfrac{\partial^2}{\partial c^2} \log p(\bm{y}, \bm{\theta})\right)^{-1} 
%     % \nonumber\\
%     % &= - \mathbb{E}_{ \Theta} \left(\mathbb{E}_{Y\vert\Theta}\left(\dfrac{\partial^2}{\partial c^2} \log p(\bm{y}, \bm{\theta})\right)\right)^{-1}
% \end{align}
% where $p(\bm{y}, \bm{\theta}) = p(\bm{y} \vert \bm{\theta})p(\bm{\theta})$. We assume that the noise signals and channels of different sampled frequencies are independent of each other, then we have
% \begin{align}
%     \! \! p(\bm{y} \vert \bm{\theta}) = \left(\dfrac{1}{\pi \sigma^2} \right)^L \! \exp \left( -\dfrac{1}{\sigma^2} \sum_{k=1}^L \left\vert y_k - h_{k}^r \gamma_k(c) h_k^t \right\vert^2\right)
% \end{align}
% and
% \begin{align}
%     p(\bm{\theta}) = p(c)\prod_{k=1}^L p(h_k^r) p(h_k^t)
% \end{align}

\begin{proposition}
    BCRB of the sensing target can be given by
    \begin{align}
        % (\text{BCRB})_{11} = \dfrac{1}{a - \bm{b}^T\bm{B}^{-1}\bm{b}} = \dfrac{1}{a - \sum_{k=1}^L\bm{b}_k^T\bm{B}_k^{-1}\bm{b}_k},
        \label{BCRB}
        (\text{BCRB})_{c} = \dfrac{1}{a - \sum_{k=1}^L\bm{b}_k^H\bm{D}_k^{-1}\bm{b}_k},
    \end{align}
    where $a$ and $D_k$ are the second-order partial derivative of $\log p(\bm{y}, \bm{\theta})$ with respect to the sensing target $c$ and the channel $\bm{h}_k$, respectively.
    % , given by
    % \begin{align}
    %     a = \mathbb{E}_{\Theta} \left( \dfrac{2}{\sigma^2}\sum_{k=1}^L \left\vert h_k^r 
    %     \gamma_k'(c) h_k^t\right\vert^2  - \dfrac{\partial^2}{\partial c^2} \log p(c)\right),
    % \end{align}
    % where $\gamma_k'(c) = \frac{\partial \gamma_k(c)}{\partial c}$. 
    $\bm{b}_k$ is the mixed second-order partial derivative with respect to sensing target $c$ and channel $\bm{h}_k$, representing the coupling between them.
    % $\bm{D}_k$ is the second-order partial derivative with respect to the channel $\bm{h}_k$.
    % , given by
    % \begin{align}
    %     \bm{D}_k =
    %     \begin{bmatrix}
    %         \bm{D}_{k1} & \bm{D}_{k3}^H \\
    %         \bm{D}_{k3} & \bm{D}_{k2}
    %     \end{bmatrix}
    % \end{align}
    The specific expressions are given in Appendix \ref{appen:A}.

    \begin{IEEEproof}
		See Appendix \ref{appen:A}.
	\end{IEEEproof}
\end{proposition}

% Rician Fading Channel

Under Rician fading, channel coefficients are modeled as
\begin{align}
    h_k^r, h_k^t \sim \mathcal{CN}\left( \sqrt{\dfrac{\kappa}{\kappa+1}}, \dfrac{1}{\kappa+1}\right), \ \  k = 1, \dots, L,
\end{align}
where $\kappa$ is the Rician factor. Then, the BCRB of the sensing target can be simplified as
\begin{align}
    \label{eq:BCRB_Rician}
    (\text{BCRB}&)_{c} = \left(\dfrac{2}{\sigma^2}\sum_{k=1}^L\mathbb{E}_{c} \left(  \left\vert \gamma_k'(c) \right\vert^2 \right)  - \mathbb{E}_{c}\left( \dfrac{\partial^2}{\partial c^2} \log p(c)\right) \right. \nonumber\\
    &\left. - \dfrac{4}{\sigma^2}\sum_{k=1}^L \dfrac{\kappa\vert \mathbb{E}_c(\overline{\gamma_k'(c)}\gamma_k(c))\vert^2}{(2\kappa+1)\mathbb{E}_c\vert \gamma_k(c)\vert^2 + \sigma^2(\kappa+1)^2} \right)^{-1}\!\!\!\!.
\end{align}

\section{BCRB Analysis}
\label{sec:BCRB_an}

Based on the derived BCRB expression (\ref{eq:BCRB_Rician}), we analyze the influence of the absorption peak shape and the number of subcarriers. We assume the sensing target follows the Gaussian distribution $c \sim \mathcal{N}(\mu_c, \sigma_c^2)$. The frequency response of the sensor is modeled by a Lorentzian absorption peak\cite{b15}, whose center frequency shifts linearly with the sensing target, i.e.,
\begin{align}
    \gamma(f,c) = 1-\dfrac{A}{1+j\dfrac{f-(\alpha c +\beta))}{\Gamma}},
\end{align}
where $2\Gamma$ is the full-width at half-maximum~(FWHM), and $A$ is the absorption depth, ranging from $0$ to $1$. In the following, we denote $(f-(\alpha c +\beta))/\Gamma = x$ for simplicity. 

% Then, the partial derivative of $\gamma(f,c)$ is calculated by
% \begin{align}
%     \gamma_k'(c)= -\dfrac{jA\alpha}{\Gamma(1+jx)^2}.
% \end{align}

% \begin{align}
%     \mathbb{E}_c\vert \gamma_k'(c)\vert^2 = \mathbb{E}_c\left( \dfrac{|A|^2 \alpha^2}{\left(\Gamma+\frac{(f_k-\alpha c-\beta)^2}{\Gamma}\right)^2}\right)
% \end{align}

% \begin{align}
%     \mathbb{E}_c\vert \gamma_k(c)\vert^2 = \mathbb{E}_c \left(1+\dfrac{A^2-A}{1+\left(\frac{f_k-\alpha c -\beta}{\Gamma}\right)^2}\right)
% \end{align}

\subsection{Single Subcarrier}

% We adopt the 3GPP model of Rician fading channel. The Rician factor can be given based on the distance between the transceiver and the sensor $d$, specifically
% \begin{align}
%     \kappa = 13-0.03\times d  \ \ [dB]
% \end{align}
% The 
First, we consider the situation of using a single subcarrier. We set the frequency as $f=\alpha\mu_c+\beta+\Delta f$, where $\Delta f$ is a small deviation from the center frequency. 
% In this case, $x$ follows the distribution $x(c)\sim \mathcal{N} \left(\Delta f/\Gamma, \alpha^2\sigma_c^2/\Gamma^2\right)$.

% \begin{align}
%     x(c)\sim \mathcal{N} \left(\frac{\Delta f}{\Gamma},\ \frac{\alpha^2\sigma_c^2}{\Gamma^2}\right).
% \end{align}
% When $f = \alpha \mu_c + \beta$, the term $A_1$ is the largest. In this subsection, we set the frequency as $f = \alpha \mu_c + \beta$. In this case, $x(c)$ follows Gaussian distribution with zero-mean.

\begin{figure*}[t!]
	\centering
	\includegraphics[scale=0.41]{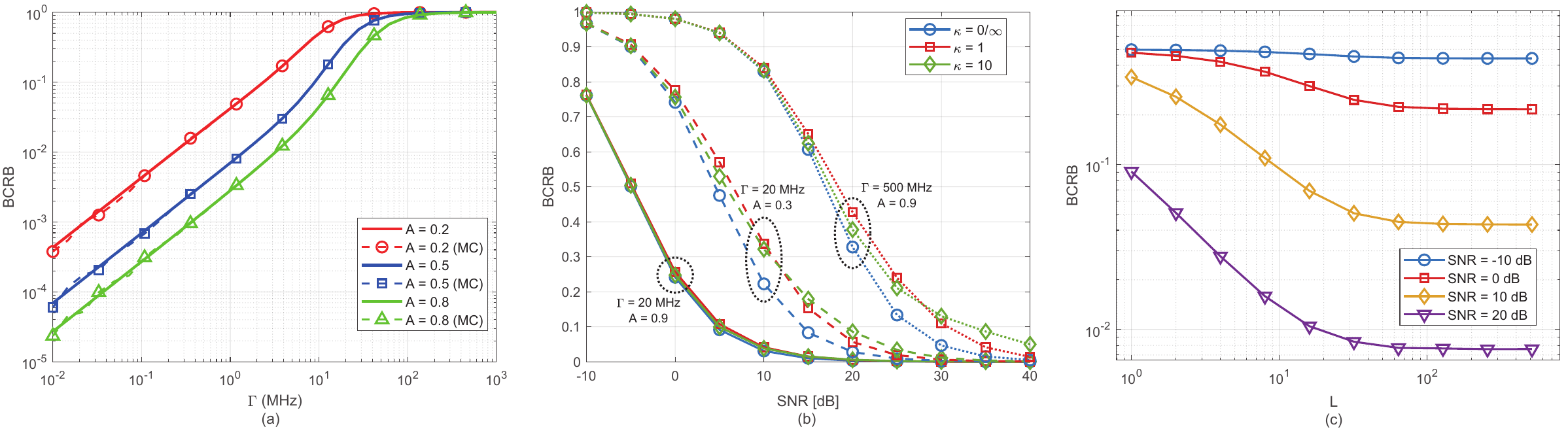}
    
	\caption{(a) Analytical and Monte Carlo results of the BCRB with respect to the FWHM and 
    depth. (b) The BCRB under different Rician factors versus the SNR. (c) BCRB versus the number of subcarriers for different SNR conditions.}
	\label{fig}
\end{figure*}

\begin{proposition}
    When $\frac{\Gamma}{\alpha \sigma_c} \gg 1$, the expectations $C_1$ and $C_3$ in $(\text{BCRB})_c$ satisfy
    \begin{align}
        C_1 &\propto \dfrac{A^2\alpha^2}{\Gamma^2},\\ 
        C_2 &\propto \dfrac{A^2\alpha^2 }{\Gamma^2}(1-A)^2.
    \end{align}
    where $C_1 = \mathbb{E}_{c} \left(  \left\vert \gamma_k'(c) \right\vert^2 \right)$, and $C_2 = \vert \mathbb{E}_c(\overline{\gamma_k'(c)}\gamma_k(c))\vert^2$. Therefore, when $A \to 1$, we have $C_1 \gg C_3$.
    \begin{IEEEproof}
		When $\frac{\Gamma}{\alpha \sigma_c} \gg 1$, the variation of $x$ is very small. Furthermore, since the $\Delta f$ is a small deviation from the center frequency, the mean of $x$ also approaches $0$. Thus, the $C_1$ and $C_3$ can be accurately approximated by evaluating their kernels at the limit $x \to 0$, yielding the desired results.
	\end{IEEEproof}
\end{proposition}

\begin{proposition}
    When $\frac{\Gamma}{\alpha \sigma_c} \ll 1$, $C_1$ and $C_3$ satisfy
    \begin{align}
        C_1 &\propto \dfrac{A^2 \alpha}{\Gamma} \phi\left(\frac{\Delta f}{\alpha \sigma_c}\right), \\ 
        C_2 &\propto A^4 \phi^2\left(\frac{\Delta f}{\alpha \sigma_c}\right),
        % \dfrac{C_1}{C_2} &\propto  \dfrac{\alpha}{\Gamma A^2} \phi^{-1}\left(\frac{\Delta f}{\alpha \sigma_c}\right)\gg 1,
    \end{align}
    where $\phi$ is the standard normal probability density function. Note that under this condition, we have $C_1 \gg C_3$.
    \begin{IEEEproof}
		See Appendix \ref{appen:C}.
	\end{IEEEproof}
\end{proposition}

% \textit{Remark 1:} Propositions 3 and 4 characterize the dominance of $A_1$ in $(\mathrm{BCRB})_c$ under two regimes. For $\Gamma/(\alpha\sigma_c)\ll 1$, the ratio $A_1/A_3\gg 1$ implies $A_3$ is negligible and $A_1$ governs sensing accuracy. Conversely, when $\Gamma/(\alpha\sigma_c)\gg 1$, $A_3$ is suppressed by $(1-A)^2$; specifically, for $A \to 1$, $A_3$ becomes vanishingly small, ensuring the first term still dominates.

% Propositions 1 and 2 characterize the dominance of $A_1$ in $(\mathrm{BCRB})_c$ under two regimes. When $\Gamma/(\alpha\sigma_c)\ll 1$, we have $A_1/A_3\gg 1$, indicating that the cross term $A_3$ is negligible and the sensing accuracy is mainly governed by $A_1$. When $\Gamma/(\alpha\sigma_c)\gg 1$, $A_1$ and $A_3$ scale in the same order with respect to $\Gamma$ and $\alpha$, while $A_3$ is additionally suppressed by $(1-A)^2$; in particular, for $A \to 1$, $A_3$ becomes small and the first term still dominates.

\textit{Remark 1:} Under the above two scenarios, $(\mathrm{BCRB})_c \approx 1/C_1$. The lower bound error is inversely proportional to the squared absorption depth. For peak shift velocity and FWHM, the error scales linearly as $\Gamma/\alpha$ when $\Gamma/(\alpha\sigma_c) \ll 1$, and quadratically as $(\Gamma/\alpha)^2$ when $\Gamma/(\alpha\sigma_c) \gg 1$.

\subsection{Multiple Subcarriers}

We consider the situation with multiple subcarriers. We assume $\Gamma \gg \delta f$, where $\delta f$ is the subcarrier spacing. When the bandwidth $B$ is much larger than $\Gamma$, we can prove that 
\begin{align}
    \mathbb{E}_{c} \left(  \sum_{k=1}^{L} \left\vert \gamma_k'(c) \right\vert^2 \right) \approx \dfrac{1}{\delta f} \mathbb{E}_c \left( \int \left\vert \dfrac{\partial \gamma(f,c)}{\partial c}\right\vert^2 df\right) = \dfrac{A^2 \alpha^2}{\Gamma \delta f} \frac{\pi}{2}.
\end{align}

\textit{Remark 2:} By approximating $(\mathrm{BCRB})_c \approx 1/C_1$, the lower error  bound scales linearly as $\Gamma/A^2\alpha^2$.

\subsection{Subcarrier Selection}

Denote the contribution of the $k$-th subcarrier as
\begin{align}
    B_k = \mathbb{E}_{c}  \left\vert \gamma_k'(c) \right\vert^2 - \dfrac{2\kappa\vert \mathbb{E}_c(\overline{\gamma_k'(c)}\gamma_k(c))\vert^2}{(2\kappa+1)\mathbb{E}_c\vert \gamma_k(c)\vert^2 + \sigma^2(\kappa+1)^2}.
\end{align}
Since the BCRB is inversely proportional to the sum of the $B_k$ terms, the optimal estimation strategy is to choose subcarriers $f_k$ that maximize their individual contribution $B_k$.

% \textit{Remark 3:}
% The best estimation strategy is to choose $f_k$ with the maximum $B_k$.

Moreover, $B_k$ can be proved to be strictly positive. Specifically, based on the Cauchy-Schwarz inequality, we have
% \begin{align}
%     \vert \mathbb{E}_c(\overline{\gamma_k'(c)}\gamma_k(c)) \vert^2 \leq \mathbb{E}_c \vert \gamma_k'(c) \vert^2 \mathbb{E}_c \vert \gamma_k(c) \vert^2.
% \end{align}
% Therefore, $B_k$ satisfies
% \begin{align}
%     % B_k &\geq \dfrac{2}{\sigma^2} \mathbb{E}_{c}  \left\vert \gamma_k'(c) \right\vert^2 - \dfrac{4}{\sigma^2} \dfrac{\kappa\mathbb{E}_c \vert \gamma_k'(c) \vert^2 \mathbb{E}_c \vert \gamma_k(c) \vert^2 }{(2\kappa+1)\mathbb{E}_c\vert \gamma_k(c)\vert^2 + \sigma^2(\kappa+1)^2} \nonumber\\
%     B_k >  \mathbb{E}_{c}  \left\vert \gamma_k'(c) \right\vert^2 - \dfrac{2\kappa\mathbb{E}_c \vert \gamma_k'(c) \vert^2 \mathbb{E}_c \vert \gamma_k(c) \vert^2 }{2\kappa\mathbb{E}_c\vert \gamma_k(c)\vert^2} = 0.
% \end{align}

\begin{align}
    B_k &\geq \mathbb{E}_{c}  \left\vert \gamma_k'(c) \right\vert^2 - \dfrac{2\kappa\mathbb{E}_c \vert \gamma_k'(c) \vert^2 \mathbb{E}_c \vert \gamma_k(c) \vert^2 }{(2\kappa+1)\mathbb{E}_c\vert \gamma_k(c)\vert^2 + \sigma^2(\kappa+1)^2} > 0.
\end{align}

\textit{Remark 3:}
The BCRB can be monotonically decreased by adding additional subcarriers.

\section{Simulation Results}
\label{sec:simu}
This section presents the simulation results of the BCRB under the Rician fading channel. We assume that the sensing target follows a real Gaussian distribution.
% The central frequency is set as $f_c = 5.8$GHz~\cite{b17}.

Fig. \ref{fig} (a) illustrates the impact of FWHM and absorption depth on the BCRB, characterizing how peak shape influences estimation accuracy. To validate the analytical expression, we also compute the BCRB via Monte Carlo~(MC) simulations. The number of subcarriers is set as $1$ due to the complexity of MC method. The MC results show excellent agreement with analytical curves, confirming the theoretical derivation. The results verify the scaling laws of BCRB relative to $\Gamma$ and $A$. 
% Specifically, BCRB increases linearly with $\Gamma$ at small $\Gamma$, and then it reflects the quadratic sensitivity before saturating at the contribution of the prior distribution. For fixed $\Gamma$, increasing $A$ from $0.2$ to $0.8$ reduces BCRB by nearly two orders of magnitude, validating $A^{-2}$ scaling.

% Moreover, as the Rician factor increases, the BCRB initially increases and then decreases, consistent with the previous theoretical analysis. In addition, we plot the asymptotic bound as $\kappa \to \infty$, which characterizes the limiting error floor for any Rician channel in such a system setup.

% \begin{figure}
%     \centering
%     \includegraphics[width=0.75\linewidth]{1_L1.pdf}
%     \caption{Analytical and Monte Carlo results of the BCRB with respect to the Rician factor.}
%     \label{fig:1}
% \end{figure}

Fig. \ref{fig} (b) shows the BCRB versus SNR, with $L = 128$ subcarriers. The results demonstrate that the BCRB decreases monotonically from the prior variance toward $0$ as the SNR increases for all Rician factors. The BCRB is lower and more robust to Rician factors for lower FWHM and higher peak absorption depth, validating the theoretical results. It is also observed that Rayleigh fading and LoS channel result in the lowest theoretical error. For Rayleigh fading, the derivation of the BCRB reveals that $\bm{b}_k = 0$ in this scenario, implying that the channel coefficients and the sensing target are statistically decoupled. Regarding the LoS channel, this occurs because no channel randomness is introduced into the system.

% exhibiting a characteristic three-stage behavior. In the very low-SNR regime, the bound is dominated by the prior distribution, maintaining a high, nearly constant value (the prior variance). As the SNR enters the intermediate range, the observation begins to contribute significantly, leading to a rapid reduction in the BCRB. Finally, at high SNR, the bound approaches zero, signifying the potential for near-perfect estimation.

% \begin{figure}
%     \centering
%     \includegraphics[width=0.75\linewidth]{2_L1.pdf}
%     \caption{The BCRB under different Rician factor versus the SNR.}
%     \label{fig:2}
% \end{figure}

% The Rician factor is set to $\kappa = 1$. 
Fig. \ref{fig} (c) shows the impact of the number of subcarriers $L$. The subcarrier spacing and the center frequency are fixed, while $L$ increases by adding subcarriers further from the center frequency. Results show that increasing $L$ initially yields a significant reduction in BCRB, as more subcarriers provide additional independent measurements of the sensing target. However, the improvement becomes marginal when $L$ is sufficiently large. This is because the additional subcarriers lie in the frequency ranges where the reflection coefficient is less sensitive to the sensing target.
% Consequently, their contribution to the overall estimation accuracy is limited.

% \begin{figure}
%     \centering
%     \includegraphics[width=0.75\linewidth]{3.pdf}
%     \caption{BCRB versus SNR for different numbers of subcarriers.}
%     \label{fig:3}
% \end{figure}

\section{Conclusion}
\label{sec:con}

In this letter, we considered a meta-backscatter sensing framework that estimates the sensing target by analyzing the frequency response of the meta-material sensor. We derived the BCRB for the sensing target and analyzed the BCRB under the Rician fading channel. Furthermore, we established the scaling law of the BCRB relative to the absorption peak parameters under a Lorentzian peak profile. The simulations demonstrated that: 1) the channel randomness does not cause the problem of error floor for the estimation of the sensing target when SNR is sufficiently large; 2) the BCRB is lower and more robust to Rician fading conditions with lower FWHM and higher absorption depth; 3) the BCRB can be monotonically decreased by adding additional subcarriers.

\begin{appendices}

\section{The Derivation of BCRB}
\label{appen:A}
% In this appendix, we derive the BCRB of the sensing target. 
As the channel coefficients and the sensing target are both unknown, we define the estimated parameter as $\bm{\theta} = [c, \bm{h}]$, and compute the Bayesian Fisher information matrix~(BFIM)
% Specifically, we have $\text{BFIM}=\bm{J}+\bm{L}$, 
% Specifically, we first calculate the Fisher information matrix~(FIM), and then take the inverse. The FIM is defined as
\begin{align}
    \!\text{BFIM} = \!-\mathbb{E}_{\Theta}\mathbb{E}_{Y \vert \Theta}\left(\dfrac{\partial^2}{\partial \bm{\theta} \partial \bm{\theta}^T}\log (p(\bm{y},\bm{\theta}))\right) = \bm{J} + \bm{U},
    % &= -\mathbb{E}_{\Theta}\mathbb{E}_{Y \vert \Theta} \left(\dfrac{\partial^2}{\partial \theta_i \partial \theta_j}(\log p(\bm{y}\vert\bm{\theta}) + \log p(\bm{\theta}))\right) \nonumber\\
\end{align}
where $\bm{J}$ is the conditional FIM derived from the conditional likelihood function $p(\bm{y}\vert \bm{\theta})$, and $\bm{U}$ is the contribution of the prior distribution derived from $p(\bm{\theta})$. The BCRB for the sensing target $c$ is then the $(1,1)$-th element of $(\text{BFIM})^{-1}$. 

Since the channel coefficients and sensing target are assumed to be independent, $\bm{U}$ is block-diagonal. The specific expression of $\bm{U}$ can be easily calculated by taking the derivative of the corresponding probability density functions. In the following, we focus on the calculation of $\bm{J}$.

% where $\bm{J}$ denotes the conditional FIM determined by the conditional likelihood function $p(\bm{y}\vert\bm{\theta})$, and $\bm{L}$ corresponds to the contribution of the prior distribution. Obviously, the FIM comprises three types of entries: the second-order derivatives with respect to the sensing target, the cross-derivative between the sensing target and the channel coefficient, and the second-order derivatives with respect to the channel coefficient vector. In the following, we calculate each type respectively.

% \subsection{Calculation of Conditional BFIM Elements $\bm{J}$}

% The elements of $\bm{J}$ are calculated using $J_{ij} = -\mathbb{E}_{\Theta}\mathbb{E}_{Y \vert \Theta}\left(\frac{\partial^2}{\partial \theta_i \partial \theta_j}\log p(\bm{y} \vert \bm{\theta})\right)$

% \subsubsection{Sensing target term}

First, we consider the sensing target term. The second-order partial derivative of $\log p(\bm{y}\vert\bm{\theta})$ over $c$ is given by
\begin{align}
\label{(1,1)}
    \dfrac{\partial^2}{\partial c^2}&\log p(\bm{y}\vert \bm{\theta})  = \dfrac{\partial^2}{\partial c^2} \left( \!- \dfrac{1}{\sigma^2}\sum_{k=1}^L \vert y_k - h_{k}^r \gamma_k(c)h^t_k\vert^2\right)\!.\!
    %  = &- \dfrac{1}{\sigma^2}\sum_{k=1}^L  \left( \left(-h_k^r\dfrac{\partial^2 \gamma_k}{\partial c^2} h_k^t\right)\overline{(y_k - h_{k}^r \gamma_k(c)h^t_k)} \right.\\
    % &\left.+ \overline{\left(-h_k^r\dfrac{\partial^2 \gamma_k}{\partial c^2} h_k^t\right)}(y_k - h_{k}^r \gamma_k(c)h^t_k) + 2 \left\vert h_k^r\dfrac{\partial \gamma_k}{\partial c} h_k^t\right\vert^2\right) \nonumber.
\end{align}
% , as shown in (\ref{(1,1)}). 
% \begin{figure*}[htbp]
% \centering
%     \begin{align}
%     \label{(1,1)}
%         \dfrac{\partial^2}{\partial c^2}&\log p(\bm{y}\vert \bm{\theta})  = \dfrac{\partial^2}{\partial c^2} \left(- \dfrac{1}{\sigma^2}\sum_{k=1}^L \vert y_k - h_{k}^r \gamma_k(c)h^t_k\vert^2\right) \nonumber\\
%         % &= \dfrac{\partial^2}{\partial c^2} \left(- \dfrac{1}{\sigma^2}\sum_{k=1}^L (y_k - h_{k}^r \gamma_k(c)h^t_k)\overline{(y_k - h_{k}^r \gamma_k(c)h^t_k)}\right) \\
%         & = - \dfrac{1}{\sigma^2}\sum_{k=1}^L  \left( \left(-h_k^r\dfrac{\partial^2 \gamma_k}{\partial c^2} h_k^t\right)\overline{(y_k - h_{k}^r \gamma_k(c)h^t_k)} + \overline{\left(-h_k^r\dfrac{\partial^2 \gamma_k}{\partial c^2} h_k^t\right)}(y_k - h_{k}^r \gamma_k(c)h^t_k) + 2 \left\vert h_k^r\dfrac{\partial \gamma_k}{\partial c} h_k^t\right\vert^2\right)
%     \end{align}
% \end{figure*}
Note that the conditional expectation $\mathbb{E}_{Y\vert \Theta}(\cdot)$ is equivalent to the expectation of $n_k$. Therefore, any term that is linear in $n_k$ vanishes after taking expectation, since $\mathbb{E}(n_k) = 0$. Similar steps will be omitted in the following.
% Therefore, any term linear in $n_k$ or $\overline{n_k}$ becomes zero in expectation.
% denotes the zero-mean noise term. Taking the conditional expectation over $\bm{y}$ given $\bm{\theta}$ eliminates the first two terms involving $n_k$, mathematically, we have
% \begin{align}
%     &\mathbb{E}_{Y\vert \Theta}
%     \left(
%     \left(-h_k^{r}\gamma_k''(c) h_k^{t}\right) \overline{y_k - h_k^r\gamma_k(c)h_k^t}
%     \right)
%     =  \\
%     &\mathbb{E}_{Y\vert \Theta}
%     \left(
%     \left(-h_k^{r}\gamma_k''(c) h_k^{t}\right) \overline{n_k}
%     \right)
%     =
%     \left(-h_k^{r}\gamma_k''(c) h_k^{t}\right)
%     \mathbb{E}\!\left[\overline{n_k}\right]
%     = 0. \nonumber
% \end{align}
Then, $a^J$ simplifies to
% Note that taking the conditional expectation of $\bm{y}$ given $\bm{\theta}$ is equivalent to taking the expectation over the noise term $n_k = y_k - h_{k}^r \gamma_k(c)h^t_k$. Therefore, the expectation of the first two terms in (\ref{(1,1)}) equals $0$, 
\begin{align}
    a^J = \dfrac{2}{\sigma^2}\mathbb{E}_{\Theta} \left( \sum_{k=1}^L \left\vert h_k^r 
        \dfrac{\partial \gamma_k(c)}{\partial c} h_k^t\right\vert^2\right).
\end{align}

% \subsubsection{Cross-terms between sensing target and the channel coefficients}

% We compute the second-order mixed partial derivative between the sensing target $c$ and the channel coefficients. 
% For illustration, we derive the term involving $c$ and $\Re(h_k^r)$. 
Then, we consider the cross-terms between sensing target and the channel coefficients. Since the channel coefficients are complex, we adopt the Wirtinger calculus\cite{b1}, which treats $h_k^r$ and $\overline{h_k^r}$ as independent variables.
Then $\bm{b}_k$ can be given by
\begin{align}
    \bm{b}_k = \dfrac{2}{\sigma^2} \mathbb{E}_{\Theta}
    \begin{bmatrix}
        \Re(\overline{h_{k}^r}\vert h_k^t\vert^2 \overline{\gamma_k'(c)} \gamma_k(c)) \\
        -\Im(\overline{h_{k}^r}\vert h_k^t\vert^2 \overline{\gamma_k'(c)} \gamma_k(c)) \\
        \Re(\vert h_k^r\vert^2\overline{h_{k}^t} \ \overline{\gamma_k'(c)} \gamma_k(c)) \\
        -\Im(\vert h_k^r\vert^2\overline{h_{k}^t} \ \overline{\gamma_k'(c)} \gamma_k(c))
    \end{bmatrix}.
\end{align}

% we calculate the $(1,i)$-th, $(i \neq 1)$ elements. We give the calculation process of $(1, 4k-2)$-th element, and the rest can be calculated in a similar manner.
% \begin{align}
%     \dfrac{\partial^2}{\partial c\partial \Re(h_k^r)} \log p(\bm{y}\vert \bm{\theta}) &= \dfrac{\partial}{\partial c} \left( \dfrac{\partial}{\partial h_k^r} +\dfrac{\partial}{\partial \overline{h_k^r}}\right) \log p(\bm{y}\vert \bm{\theta}) \\
%     \dfrac{\partial^2}{\partial c\partial \Im(h_k^r)} \log p(\bm{y}\vert \bm{\theta}) &= i\dfrac{\partial}{\partial c} \left( \dfrac{\partial}{\partial h_k^r} -\dfrac{\partial}{\partial \overline{h_k^r}}\right) \log p(\bm{y}\vert \bm{\theta})
% \end{align}

% \subsubsection{Channel terms}
Finally, we calculate the channel terms. Since the channels of different subcarriers are assumed independent, the cross-terms between different subcarriers are all $0$.
% i.e., $f_k\neq f_{k'}$,
% First, we consider the cross-terms between different subcarriers, i.e., $k\neq k'$. The second-order partial derivative of $h_{k}^r$ and $h_{k'}^r$ can be calculated as below.
% \begin{align}
%     \dfrac{\partial^2}{\partial h_k^r\partial h_{k'}^r}& \log  p(\bm{y}\vert \bm{\theta}) \nonumber\\
%      = &-\dfrac{1}{\sigma^2} \dfrac{\partial}{\partial h_{k'}^r}\left( \left(-\gamma_k h_k^t\right)\overline{(y_k - h_{k}^r \gamma_k(c)h^t_k)} \right. \nonumber\\
%     & \left. + (y_k - h_{k}^r \gamma_k(c)h^t_k)\overline{\left(-\gamma_k h_k^t\right)}\right) = 0.
% \end{align}
% Similarly, it can be found that all cross-terms between different $k$ are $0$. 
The second-order partial derivatives with respect to $\bm{h}_k^r$ and $\bm{h}_k^t$ are given by
\begin{align}
    \bm{D}_{k1}^J = \frac{2}{\sigma^2}\mathbb{E}_{\Theta}\left(\vert \gamma_k(c)\vert^2 \vert h_k^t\vert^2 \right) \bm{I}_2, \\
    \bm{D}_{k2}^J = \frac{2}{\sigma^2}\mathbb{E}_{\Theta}\left(\vert \gamma_k(c)\vert^2 \vert h_k^r\vert^2 \right) \bm{I}_2.
\end{align}
% where $\bm{I}_2$ is a $2\times 2$ identity matrix. 
% Similarly, the second-order partial derivatives with respect to $\bm{h}_k^t = [\Re(h_k^t),\Im(h_k^t)]$ can be given by
% \begin{align}
%     \bm{D}_{k2}^J = \frac{2}{\sigma^2}\mathbb{E}_{\Theta}\left(\vert \gamma_k(c)\vert^2 \vert h_k^r\vert^2 \right) \bm{I}_2.
% \end{align}
% The second-order partial derivatives with respect to $\bm{h}_k^t = [\Re(h_k^t),\Im(h_k^t)]$ can be obtained similarly.

The cross-terms of the second-order partial derivatives for $h_k^r$ and $h_k^t$ are expressed as follows.
% \begin{align}
%     \dfrac{\partial^2}{\partial \Re(h_k^r)\partial \Re(h_k^t)} \log p(\bm{y}\vert \bm{\theta}) = \frac{1}{\sigma^2}\left(2\Re(h_k^r\overline{h_k^t}\vert \gamma_k(c)\vert^2 ) \right.\nonumber\\
%     \left. - \gamma_k \overline{(y_k - h_k^r\gamma_kh_k^t)} - \overline{\gamma_k}(y_k - h_k^r\gamma_kh_k^t)\right).
% \end{align}
% After taking the expectation, we have
% \begin{align}
%     \mathbb{E}_{Y, \Theta}\left(\dfrac{\partial^2 \log p(\bm{y}\vert \bm{\theta})}{\partial \Re(h_k^r)\partial \Re(h_k^t)}  \right) = \frac{2}{\sigma^2}\Re(h_k^r\overline{h_k^t}\vert \gamma_k(c)\vert^2 ).
% \end{align}
% The rest terms are calculated similarly, summarized as below.
\begin{align}
    \!\!\bm{D}_{k3} \! = \!\frac{2}{\sigma^2}\mathbb{E}_{\Theta} \!\!
    \begin{bmatrix}
        \Re(h_k^r\overline{h_k^t}\vert \gamma_k(c)\vert^2 ) \!\!& -\Im(h_k^r\overline{h_k^t}\vert \gamma_k(c)\vert^2 ) \\
        -\Im(h_k^r\overline{h_k^t}\vert \gamma_k(c)\vert^2 ) \!\!& \Re(h_k^r\overline{h_k^t}\vert \gamma_k(c)\vert^2 )
    \end{bmatrix}\!. \!\!
\end{align}

% \subsection{Calculation of Prior Information $\bm{U}$}

% The element associated with $c$ and $h_k^r$ are given by
% \begin{align}
%     a^U &= -\mathbb{E}_{c}\left( \dfrac{\partial^2}{\partial c^2}\log p(c)\right), \\
%     \bm{D}_{k1}^U &= - \mathbb{E}_{H} \left(\dfrac{\partial^2 \log p(h_k^r)}{\partial \bm{h}_k^r \partial {\bm{h}_k^r}^T}  \right),
% \end{align}
% The channel prior block corresponding to $h_k^r$ can be given by
% \begin{align}
%     \bm{D}_{k1}^U = - \mathbb{E}_{H} \left(\dfrac{\partial^2 \log p(h_k^r)}{\partial \bm{h}_k^r \partial {\bm{h}_k^r}^T}  \right),
% \end{align}
% \begin{align}
%     \bm{D}_{k1}^U = - \mathbb{E}_{H}
%     \begin{bmatrix}
%         \dfrac{\partial^2 \log p(h_k^r)}{\partial (\Re(h_k^r))^2} & \dfrac{\partial^2 \log p(h_k^r)}{\partial \Re(h_k^r)\Im(h_k^r)} \\
%         \dfrac{\partial^2 \log p(h_k^r)}{\partial \Im(h_k^r)\Re(h_k^r)} & \dfrac{\partial^2 \log p(h_k^r)}{\partial (\Im(h_k^r))^2}
%     \end{bmatrix}
% \end{align}
% and the block $\bm{D}_{k2}^U$ corresponding to $h_k^t$ is obtained similarly.

% Since we assume that the sensing target, the channels of different subcarriers are independent, $\bm{L}$ is a block-diagonal matrix. 

% \subsection{BCRB derivation}
Collecting all the derived terms, BFIM is constructed by
% We define the aggregated elements as $a = a^J + a^U$, $\bm{D}_{ki} = \bm{D}_{k}^J + \bm{D}_{ki}^U, i= 1, 2$. The BFIM is given by the block matrix
% from the conditional FIM $\bm{J}$ and the prior matrix $\bm{U}$
\begin{align}
    \text{BFIM} = 
    \begin{bmatrix}
        a & \bm{b}^T \\
        \bm{b} & \bm{D} 
    \end{bmatrix} \in \mathbb{R}^{(1+4L)\times (1+4L)},
\end{align}
where $a = a^J + a^U$, $\bm{b} = [\bm{b}_1 ;\bm{b}_2;...;\bm{b}_L]$, and $\bm{D} = \text{blkdiag}\{\bm{D}_1, ..., \bm{D}_L\}$, 
where $\bm{D}_k = [\bm{D}_{k1}, \bm{D}_{k3}^T;  \bm{D}_{k3}, \bm{D}_{k2}]$, $\bm{D}_{ki} = \bm{D}_{k}^J + \bm{D}_{ki}^U, i= 1, 2$. 
% By taking the inverse of BFIM, we can obtain the BCRB of the sensing target as
% mathematically expressed as
% \begin{align}
%     (\text{BCRB})_{c} = \dfrac{1}{a - \bm{b}^T\bm{D}^{-1}\bm{b}}.
% \end{align}
Since $\bm{D}$ is a block-diagonal matrix, the $(1,1)$-th element of the inverse of BFIM simplifies to (\ref{BCRB}).
% \begin{align}
%     (\text{BCRB})_{c} =
%     \dfrac{1}{a - \sum_{k=1}^L \bm{b}_k^T\bm{D}_k^{-1}\bm{b}_k}.
% \end{align}
Hence, the proof of Proposition 1 is completed.

\section{Proof of Proposition 3}
\label{appen:C}

We first derive the scaling of $C_1=\mathbb{E}_c|\gamma'(c)|^2$ in the regime $\Gamma/(\alpha\sigma_c)\ll 1$. Specially,
% \begin{align}
%     \vert \gamma(c)'\vert^2 = \dfrac{A^2 \alpha^2}{\Gamma^2 (1+x^2)^2}
% \end{align}
% The derivative of $\gamma_k(c)$ is expressed as
% \begin{align}
%     \gamma_k'(c)= -\dfrac{jA\alpha}{\Gamma(1+jx)^2}.
% \end{align}
% After taking the expectation, we have
\begin{align}
    \mathbb{E}_c\vert \gamma(c)'\vert^2 = \dfrac{A^2 \alpha^2}{\Gamma^2} \mathbb{E}_x \left(\dfrac{1}{(1+x^2)^2} \right).
\end{align}
Let $x = s u$ with $u \sim \mathcal{N}(r, 1)$, where we denote the constants $s = \frac{\alpha \sigma_c}{\Gamma}$ and $r = \frac{\Delta f}{\alpha \sigma_c}$. Then, we have
% \begin{align}
%     \mathbb{E}_x \left(\dfrac{1}{(1+x^2)^2} \right) &= \mathbb{E}_u \left(\dfrac{1}{(1+s^2u^2)^2} \right) \nonumber\\
%     &= \int  \phi(u-r) \dfrac{1}{(1+s^2 u^2)^2}du  \\
%     &= \frac{1}{s} \int \phi\left(\frac{v}{s}-r \right) \dfrac{1}{(1+v^2)^2}dv. \nonumber
% \end{align}
\begin{align}
    \mathbb{E}_x \left(\dfrac{1}{(1+x^2)^2} \right) &= \frac{1}{s} \int \phi\left(\frac{v}{s}-r \right) \dfrac{1}{(1+v^2)^2}dv.
\end{align}

Based on dominated convergence theorem\cite{b20}, we have
\begin{align}
    \!\int \!\!\phi \left(\frac{v}{s}-r \right) \dfrac{1}{(1+v^2)^2}dv
    \xrightarrow{s \to \infty} \phi(r) \int \dfrac{1}{(1+x^2)^2}dx. \!
\end{align}

Therefore, $C_1$ can be expressed as
\begin{align}
    \mathbb{E}_c\vert \gamma(c)'\vert^2 \to \frac{1}{s} \dfrac{\pi}{2} \phi(r) \dfrac{A^2 \alpha^2}{\Gamma^2} = \dfrac{\pi}{2} \phi\left(\frac{\Delta f}{\alpha \sigma_c}\right)\dfrac{A^2 \alpha}{\Gamma\sigma_c}.
\end{align}

% Then we calculate the term $C_2$, which can be expressed as
% \begin{align}
%     \overline{\gamma_k'(c)}\gamma_k(c) = j\dfrac{A \alpha}{\Gamma} \dfrac{-x^2 + jx(2-A) + 1-A}{(1+x^2)^2}.
% \end{align}
In a similar way, $C_2$ can be calculated by
\begin{align}
    \vert \mathbb{E}_c(\overline{\gamma_k'(c)}\gamma_k(c))\vert^2 \to  \dfrac{\pi^2}{4} \phi^2\left(\frac{\Delta f}{\alpha \sigma_c}\right) \dfrac{A^4 }{\sigma_c^2}.
\end{align}

Combining the above scalings, we obtain $\frac{C_1}{C_2}\gg 1$.
% \begin{align}
%     \dfrac{C_1}{C_2} =\dfrac{2}{\pi} \dfrac{\sigma_c \alpha}{\Gamma A^2} \phi^{-1}\left(\frac{\Delta f}{\alpha \sigma_c}\right) \gg 1.
% \end{align}
Hence, the proof of Proposition 3 is completed.

% In a similar way, we can calculate
% \begin{align}
%     \mathbb{E}\left(\dfrac{-x^2 + 1-A}{(1+x^2)^2} \right) &\to -\frac{1}{s} \dfrac{\pi}{2} \phi(r) A, \\
%     \mathbb{E} \left(\dfrac{x}{(1+x^2)^2} \right) &\to \frac{r}{s^2} \dfrac{\pi}{2} \phi(r).
% \end{align}

% since it is an odd function.

% \begin{align}
%     \vert \mathbb{E}_c(\overline{\gamma_k'(c)}\gamma_k(c))\vert^2 \to \frac{1}{s^2} \dfrac{\pi^2}{4} \phi^2(r) \dfrac{A^4 \alpha^2}{\Gamma^2} = \dfrac{\pi^2}{4} \phi^2\left(\frac{\Delta f}{\alpha \sigma_c}\right) \dfrac{A^4 }{\sigma_c^2}
% \end{align}

\end{appendices}

\bibliographystyle{IEEEtran}
\bibliography{ref}

\end{document}